# Generative Artificial Intelligence for Academic Research: Evidence from Guidance Issued for Researchers by Higher Education Institutions in the United States

Amrita Ganguly, Aditya Johri[1], Areej Ali, and Nora McDonald
*George Mason University*

**Abstract**
The recent development and use of generative AI (GenAI) has signaled a significant shift in research activities such as brainstorming, proposal writing, dissemination, and even reviewing, resulting in questions about how to balance the seemingly productive uses of GenAI with ethical concerns such as authorship and copyright issues, use biased training data, lack of transparency, and impact on user privacy. To address these concerns, many Higher Education Institutions (HEIs) have released institutional guidance for researchers. To better understand the guidance that is being provided we report findings from a thematic analysis of guidelines from thirty HEIs in the United States that are classified as R1 or "very high research activity." We found that guidance provided to researchers: 1) asks them to refer to external sources of information such as funding agencies and publishers to keep updated and use institutional resources for training and education; 2) asks them to understand and learn about specific GenAI attributes that shape research such as predictive modeling, knowledge cutoff date, data provenance, and model limitations, and about ethical concerns such as authorship, attribution, privacy, and intellectual property issues; 3) includes instructions on how to acknowledge sources and disclose the use of GenAI, and how to communicate effectively about their GenAI use, and alerts researchers to long term implications such as over reliance on GenAI, legal consequences, and risks to their institutions from GenAI use. Overall, guidance places the onus of compliance on individual researchers making them accountable for any lapses, thereby increasing their responsibility.

## 1 Introduction

As the use of generative artificial intelligence (GenAI) increases across all facets of society, one area of significant impact is higher education institutions (HEIs). Although the initial scholarship on the use of GenAI within HEIs has focused on teaching and learning (McDonald et al., 2025; Ali et al., 2025) increasingly, studies are starting to examine how academic research is being impacted by GenAI (Abernethy, 2024; Lehr, et al., 2024; Lin, 2024; Liu and Jagadish, 2024; Godwin et al., 2024) This shift is in keeping with increased uptake of the use of GenAI for research. GenAI has many potential benefits for researchers across different stages of the research process such as data analysis, creation of content for research dissemination, and as a tool to brainstorm new ideas (Joosten et al., 2024) For instance, Delios et al. (2024) report that almost 30% of scientists are using GenAI as partners in their tasks related to research such as summarizing literature review, data analysis, grant writing and assisting with other aspects of manuscript preparation (Morocco-Clarke et al., 2024; Xames and Shefa, 2023). In a 2023 Nature survey of 1600 scientists, 30% acknowledged that they used GenAI to write academic papers, conduct literature reviews, and/or develop grant applications (Chawla, 2024). Another survey conducted by Nature among 3,838 postdocs indicated a similar level of engagement with GenAI, particularly chatbots, with 31% of respondents reporting using chatbots (Nordling, 2023). One application of GenAI in particular, Large Language Models (LLMs) based applications such as ChatGPT, has seen very high uptake as they can assist with writing which is a component of different parts of the research process (Biswas, 2023; Formosa et al., 2024). Writing was already a task that was often undertaken with the help of tools such as Grammarly, Zotero, and Evernote, among others that helped improve grammar and sentence structure and assisted with citations (Alkhaqani, 2023). The use of LLMs has now allowed researchers to use a single application for multiple writing related tasks in conjunction with functions such as data exploration and analysis (Abdelhafiz et al., 2024).

---
[1] Corresponding author: johri@gmu.edu

Although the use of GenAI for research is on the rise, the advantages of the technology are unclear and there is ambiguity about its potential benefits and as a consequence researchers are engaging with it cautiously (Nordling, 2023). One area of concern with the use of GenAI for research is how GenAI systems are developed. They are predictive models trained on extremely larger datasets and their output replicates and perpetuates any biases that exist in the training data. The initial models in particular had no access to external data to verify their outputs although this is changing with newer models post-GPT4 that have access to web searches and databases to retrieve verifiable sources. Still, the model itself still has no "knowledge" beyond the statistical regularities of its training data and the output is questionable. Van Noorden & Perkel (2023) found that while a large minority of researchers engaged with AI frequently and recognized benefits such as efficiency gains and improved accessibility for non-native English speakers, they expressed concerns about misinformation, largely due to inherent biased datasets used to train the LLM models (Yusuf et al., 2024; Biswas, 2023; Rowland, 2023). This concern with bias and false output is compounded by the fact that the output of GenAI applications looks quite coherent and plausible, leading to a lack of trust. Improper use of LLMs has also resulted in articles with hallucinations, that is, made up text such as fictional references and other factual inaccuracies. There are also concerns related to research data integrity and ownership of the generated contents (Morocco-Clarke et al., 2024; Xames and Shefa, 2023). Therefore, validation of the output is critical and the researcher has to take responsibility for the use of GenAI. Data privacy and data protection are other concerns and it is important that researchers do not assume that any information they input or share with a GenAI tool is private or secure. There are many potential risks associated with inputting sensitive, private, confidential, or proprietary data into these tools, even when a university has a license for its use, and in addition to intellectual property issues, use of certain data might violate legal or contractual requirements, in addition to expectations for privacy (Al-kfairy et al., 2024). Finally, there are concerns both with authorship and peer review of scientific work as GenAI can produce high quality articles and abstracts which were hard to distinguish from human-authored text (Khlaif et al., 2023) and is abused by many scholars to produce reviews that are generic and lack specific feedback (Liang et al., 2024, pg. 10).

## 2 Frameworks to Guide Use of GenAI for Academic Research

As a consequence of the potential concerns with GenAI use, there is an acknowledgement within research communities that guardrails for research need to be developed and followed (Delios et al., 2024; Duah & McGivern, 2024) and academic researchers are being encouraged to be cautious while using AI generated content and evaluate the output quality and accuracy (Godwin et al., 2024; Frank et al., 2024; Rowland, 2023). Initial guidance has come from HEIs, funding agencies, policy makers, and publishers, and stresses that GenAI should be seen as an assistant rather than a replacement for human effort such as critical thinking, exploratory analysis, and writing skills (Xames and Shefa, 2023; Kiley, 2024; Harding and Boyd, 2024); many publications clearly state that GenAI tools should not be listed as authors in scientific publication (Hsu, 2023; Abdelhafiz et al., 2024). As part of guidance being provided about the use of GenAI for research, certain frameworks have been advanced to map and understand the landscape. In particular, Smith et al. (2024) have advanced a strategic framework that an institution can use to map the stakeholders and activities and support responsible use of GenAI. This framework provides a comprehensive and systematic method to understand all the factors involved in the implementation of GenAI for research. Their framework is structured in four layers starting with *Context* at the top, followed by *Development* and *Implementation* layers, and ending with *Review* at the bottom. In between are *Development* and then *Implementation* layers. According to Smith et al., this is also the order in which different elements of the framework should generally be considered and implemented. The Context layer describes the external and internal policy environment that governs research integrity and research conduct and helps shape institutional responses to opportunities and risks posed by GenAI in research. The Development layer emphasizes the importance of developing a position statement to apply the principles of research integrity to the specific opportunities and challenges posed by GenAI. Implementation describes a plan to put the position statement into practice with appropriate support, processes and infrastructure. Finally, Review describes a plan to iteratively evaluate the framework to test its effectiveness and undertake revisions or updates to ensure currency. In our work, we were guided to understand the overall landscape through this framework and to examine what factors affect researchers directly and are covered by the guidance issued by institutions.

Another relevant framework advanced by Al-kfairy et al. (2024) is based on a review of 37 articles that focused specifically on the use of GenAI for research. This paper identifies eight concerns, each with implication for the use of GenAI in research: Authorship and Academic Integrity; Intellectual Property and Copyright; Privacy, Trust, and Bias; Misinformation and Deepfakes; Educational Ethics; Transparency and Accountability; Authenticity and Attribution; and Social and Economic Impact. Al-kfairy et al. (2024) explain that according to the articles that were reviewed, misgivings about who has actually written a text, a human or AI, is a concern for academic integrity reasons and even questions of attribution. Similarly, production of text or visual by AI challenges notions of copyright and intellectual property as it is unclear what the new content was derived from and how to attribute the role of a machine in the creation. Data privacy, especially if personally identifiable data is entered into a GenAI application, remains a constant risk and so does the issue of lack of transparency around data training. The reviewed articles also raised concerns with the output of GenAI systems, especially misinformation and deliberate misuse to create deepfakes, risking privacy and identify theft. The impact of GenAI on education was another specific domain related theme that emerged from the analysis raising an alert not only for increased plagiarism but decline in critical thinking and problem-solving skills. Lack of transparency due to algorithmic opacity was another concern that was addressed by the papers as a lack of transparency not only results in systemic bias and increase in discrimination, it also effects accountability and biases can go unchecked, reinforcing existing inequalities. Finally, a broader concern that emerged from the review focused on GenAI's ability to alter the landscape of work and labor, shape public discourse, and lead to the creation of regulations and law.

Finally, Lin (2024), in a recent paper in this journal, argues that there is a need to bridge the gap between abstract principles related to GenAI use of research and the everyday practices of researchers. He outlines a user-centered realistic approach with five specific goals for ethical AI use that includes developing and understanding model training and output; respecting privacy, confidentiality, and copyright issues; avoiding plagiarism and policy violations; applying AI beneficially compared to alternatives; and using AI transparently and reproducibly. Similar to Smith et al, (2024), Lin (2024) also argues the creation of documentation guidelines and development of training programs.

## 3 Research Goal and Approach

Overall, guidance for the use of GenAI in research varies greatly across journals, funding agencies, and professional associations, and also changes frequently. Although policies regarding the use of GenAI continue to evolve, it is important to continue to create some understanding of the terrain so that it is easier to see what is changing over time. Our goal was to contribute to this dialogue and add to a recent recommendation by Lin (2024), in this journal, to focus on the practical aspects of GenAI use rather than conduct a theoretical exercise on what makes for ideal research. Our work reports on what is the practical guidance being provided to researchers, how comprehensive it is, and what it means for research conduct. We take an inductive approach where we collected and analyzed the data to see what themes emerge but we were guided in this process by the frameworks advanced by the Smith et al. (2024), Al-kfairy et al. (2024), and Lin (2024). The research questions guiding our study were: 1) what institutional guidance has been provided to researchers in relation to the use of GenAI for research and 2) what attributes specific to GenAI applications have been considered in the guidance. In this paper we take the approach of forming a better understanding through the analysis of research policies advanced by HEIs.

## 4 Data and Methodology

*Data Selection*

To understand how institutions respond to GenAI use and the guidance they are proving, we analyzed the policies or guidelines they have publicly released, an approach similar to what has been used in related work (Ali, et al., 2025; de-Lima-Santos et al., 2024; McDonald et al., 2025). To assemble a corpus of institutions whose GenAI research guidelines and policies we could analyze, we focused on institutions in the United States that were designated as conducting a very high level of research based on the Carnegie Classification[2] for HEIs in the U.S. We have adopted this approach of using data from a single country and focusing on one kind of institution for consistency. The research infrastructure available to scientists and scholars, including support from the office of research is similar in these institutions. By focusing on research intensive R1 institutions, we hypothesized that we were more likely to find research policy guidelines as most R1 institutions have research offices to manage external funding and comply with federal law, and to approve research projects, including consent procedures for human subjects research. We recognize the limitations of this approach in that because R1 universities tend to be well-funded and often resource-rich in the research arena, it limits the generalization of our findings to institutions that are less resourced with lower levels of research activities, i.e. those with less advanced infrastructure and policy-making capacity. A similar approach has been taken by other scholars that study HEIs (e.g., Brown & Klein, 2020) and by us in our recently published studies where we have analyzed GenAI guidelines focused on teaching and learning (*self-references blinded for review*).

*Data Collection*

---

[2] [n. d.]. Carnegie Classification of Institutions of Higher Education®. https://carnegieclassifications.acenet.edu/.

Since there is no database containing publicly accessible policies related to the use of GenAI in research within HEIs, we create a dataset using web search. This specific research study is part of a larger project and through that work *(self-references blinded for review)* we experimented with several ways to both find all the policies that existed but also limit what we found to HEIs policies and guidelines rather than other forms of documents. Our initial keyword resulted primarily in a broad array of policies and we divided them into two sub-areas, those related to teaching and learning and those related to research. We further sub-divided these into guidelines that were specific to GenAI use with a course, i.e. found in syllabi, and those that were about teaching and learning at the institutional level (*self-references blinded for review*). Within research, we further sub-divided the policies retrieved across those relevant to research at the institution level, often but not exclusively released by the office of research, and those related to doing research for an essay or paper released by university libraries and targeted at students. In this paper we report on guidelines that referred to research at the institutional level conducted by faculty and other researchers. Our search was further complicated by different departments or offices that released these guidelines and we were inclusive in terms of including all of them in our sample. Finally, our search phrases returned documents labeled GenAI 'policies', 'guidelines', and 'best practices'. We have included all three in the paper and use the term "guidelines" to refer to them. The final dataset was created using the following keyword phrases on Google as we found these to be the most suitable for answering our research questions:

- "X University generative AI research policy"
- "X University generative AI and research policy"
- "X University generative AI in research policy office of research"

The 2021 Carnegie framework classifies 146 doctoral-granting universities in the U.S. with high research activity or R1 and through our search we found thirty policies and/or guidelines that fit our criteria. We downloaded both a PDF copy and a direct URL of *publicly accessible guidelines*. We excluded policies and/or guidelines that were available through institutional resources such as non-public SharePoint sites and would not constitute a public resource. Data was collected from July 20, 2024, to August 27, 2024. In our final dataset we included research policies which were specifically provided by research offices, offices of compliance, and offices of the provost, and other offices or departments related to research. We excluded policies and/or guidelines that vaguely mentioned research data or were not solely research-specific, as well as instructional design-related sources that were primarily about literature review or background research. Given that this is a fast-developing area, we recognize the limitation that possibly more policies have been released since we collected the data or have been revised. Our data captures a slice in time.

*Data Analysis*

Overall, we followed the process for qualitative content analysis as suggested by Zhang and Wildemuth (2009). Three researchers reviewed the data (N=30 institutions) and discussed which units/sections of the guidelines to focus on (Step 2, Fig. 1), this was guided in part by prior work, especially Smith et al. (2024) and Al-kfairy et al. (2024). For example, we decided to look for policies related to data handling, funding and grants, regulations by federal and funding agencies, human subject data considerations, etc. After familiarizing the data, three researchers did an open coding which resulted in a codebook with six categories each containing multiple codes and their definitions (Step 3, Fig. 1). For instance, the category "Data Guidance with GenAI Tools" contains two codes "GenAI Output Considerations" and "GenAI Input

Considerations" and each of these codes includes multiple subcodes. See Tables 1 and 2 for the code details. Two researchers then initiated the formal coding process and refined the codes in an inductive manner (Step 4, Fig. 1). After discussing the codes, reconciling differences, and further defining the code definitions, the researchers reached a near perfect agreement to gain inter-rater reliability (IRR) (following the process suggested in McDonald et al., 2019). In the codebook, for each university, the codes were captured only once and researchers ensured there is no duplication of a single code. Codes were further revised after completing 50% of the dataset and then finalized using a summative check (Step 4, Fig. 1). The same process was followed to capture the subcodes under the codes. Figure 2 shows the inductive framework of the code generation starting from categories to codes and subcodes.

At the end of coding process, researchers compared their results and discussed the differences to finalize the codebook. If an institution was found to have guidelines collected from multiple offices or departments, they were then treated as a collective data unit as the coding is specific to each institution. To analyze the results, researchers first counted the frequency of each code in the entire codebook (Step 6.1, Fig. 1). The primary target was to analyze the overlaps between the codes. These frequency counts (Table 1 and Table 2) were further used to determine the counts of overlapping codes and interpret the results through visualization and discussion (Step 6.1, Fig. 1).

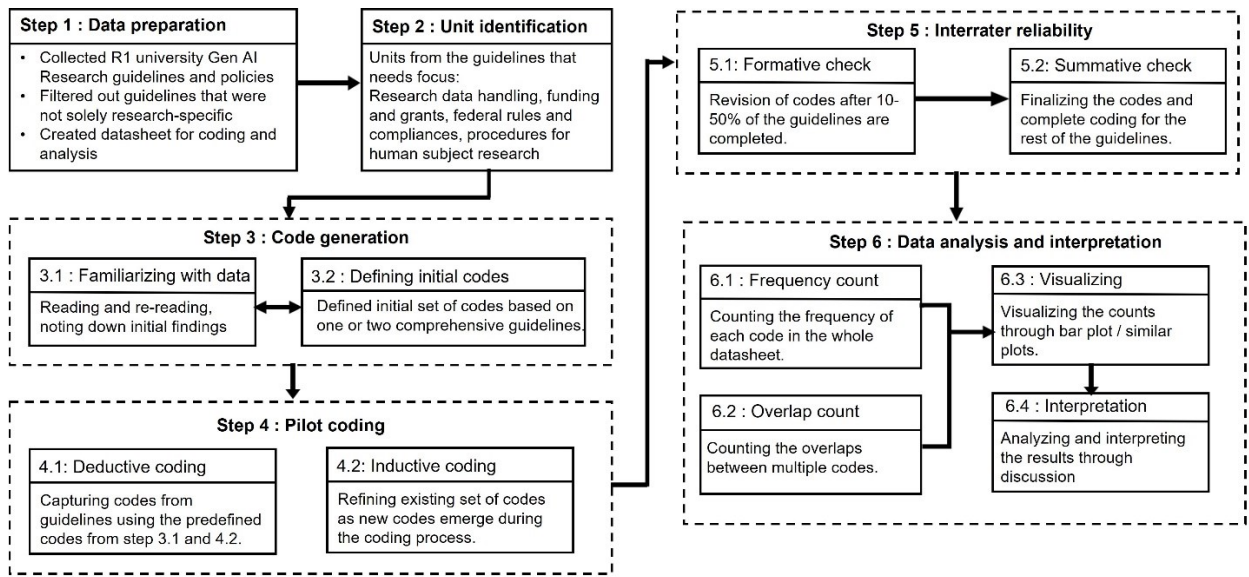

*Figure 1: A research framework outlining the steps for data coding and analysis, following Zhang and Wildermuth's content analysis approach*

Table 1 and Table 2 below list all the codes and corresponding categories for data analysis and reporting. When reporting the occurrence of each code in Table 1 and 2, percentages are out of N=30. A policy can capture multiple codes and subcodes for any given category and thus are reported out of 30 and not the total N for those codes, meaning these are not mutually exclusive. The same applies to any subcodes, they are reported out of the respective general code N. The category "GenAI Permissible Use" is an exception as it has two mutually exclusive subcodes: "Guided Use of GenAI" and "Disallowed Use of GenAI." In Table 3 below, the numbers in square brackets accompanying the institution name refer to the table in the Appendix A that lists all institutions in the dataset.

| Category | Code | Definition | N= |
|---|---|---|---|
| GenAI Permissible Use | Guided Use of GenAI | This is when policies or guidelines leave it open-ended to researchers whether they choose to use GenAI. If they choose to use GenAI in their research, they are provided specific guidelines either cautioning them, focusing on the benefits, specifying restrictions, or all aforementioned. | 28 |
| | Disallowed Use of GenAI | Provides examples when incorporating GenAI in the research process would not be allowed. Includes keywords such as "disallowed," "not acceptable," "do not," etc. | 2 |
| Data Guidance with the Use of GenAI Tools | GenAI Output Considerations | Mention risks or limitations associated with using GenAI output in research. This may include plagiarism, bias, inaccuracy of data, GenAI hallucination, etc. | 27 |
| | GenAI Input Considerations | Mentions the types of research data that should or should not be entered into GenAI tools. Types of data that can be submitted to GenAI tools. Do not enter sensitive data into public GenAI tools (e.g., PII, PHI, confidential, proprietary, etc.). References HIPAA, FERPA, or other data governance regulations. | 26 |
| Long-Term Implications of GenAI in Research | Reliance on GenAI | Refrain from relying on GenAI for decision-making or a call to action to verify output, and the need for human oversight, or overreliance may lead to the erosion of skills. | 25 |
| | Legal Implications with the use of GenAI | Agreeing to "Terms and Conditions," liability to person or organization, accusations of research misconduct, ownership of IP/IP rights, copyright issues, patent laws, third-party risks, etc. | 23 |
| Regulatory Guidance and Standards for GenAI Use | References Funding Agency Guidelines and/or Policies for the Use of GenAI | Redirects to or references official guidance/policies published by Funding Agencies, such as NSF, NIH, DOE, etc. Specifically, these guidelines/policies discuss using GenAI for writing or reviewing grants, whether it is allowed or not, the scope of limitations, etc. References can be general and non-specific. | 24 |
| | References Research Publisher Guidelines and/or Policies for the Use of GenAI | Redirects to or references official guidance/policies provided by Journals and/or Publishers, such as Wiley, Springer, Elsevier, etc. This is specifically when submitting papers/research for publication. References can be general and non-specific. | 23 |
| | Reference to Government Agency Guidelines and/or Policies for the Use of GenAI | Redirects to or references official guidance/policies published by Federal Government Agencies/Offices, such as NIST, The White House, etc, as well as State-Level Executive Orders, etc. Specifically, this is more general guidance about the responsible use of GenAI, its implications, etc. References can be general and non-specific. | 9 |
| Responsible Use of GenAI | Disclosure of GenAI | Formal or informal citation of GenAI, acknowledgment, authorship, etc. | 23 |
| | Responsible Party for GenAI Usage | Specifies who is the responsible party for adhering to these guidelines/policies (e.g., the PI, co-investigators, researchers, co-authors, scholars, etc.). It has to state a level of responsibility or accountability directed towards an individual or broader group. | 20 |
| | Communicating Responsible Use of GenAI | The Researcher or Responsible Party communicates best practices, ethics, approved uses, benefits, and risks to research staff, including subcontractors. | 9 |
| Institutional Guidance and Resources | Education and Awareness for GenAI Use | Mention training resources, workshops, updated information or other materials to train faculty and staff on GenAI best practices, implications, benefits, risks, regulations, etc. | 9 |
| | GenAI Tool Institutional Acquisition Process | Recommends or references institutions' researchers to obtain approval before procuring GenAI tools for their research. This ensures that the tools meet certain requirements. | 8 |

**Table 1: Category of Codes and Definitions for Analyzing GenAI Research Policies and/or Guidelines**

| Code | Subcode | Subbcode Definition | N= |
|---|---|---|---|
| GenAI Output Considerations | Biased Output | Based on the quality of training data, GenAI tools may produce outputs with various bias. | 23 |
| | Fabricated Output | Refers to fabricated data or non-existent information presented as factual. | 16 |
| | Inaccurate Output | Refers to misleading or incorrect output presented as factual. | 27 |
| | Knowledge Cutoff Date | Pertains to GenAI tools trained on historical data that can present data that is not current. | 7 |
| | Plagiarism | Specific to GenAI output that includes material from unacknowledged sources, presenting an ethical research violation. | 24 |
| | Other | Types of data not captured in the GenAI Output Considerations subcodes listed above. | 11 |
| GenAI Input Considerations | Operational Data | Related to operations/administrative functions in higher education, such as financial operations, university operations, etc. | 8 |
| | Human Subjects Data | Specific to research studies focusing on human data. | 10 |
| | PII | Pertains to personally identifiable information, including but not limited to full names, email addresses, phone numbers, physical addresses, dates of birth, social security numbers, etc. This may also include protected health information (PHI). | 18 |
| | Proprietary Data | Specific to non-public innovative/creative data that may be commercialized. | 12 |
| | Unpublished Research | Refers to non-public research data, draft papers, grant applications, proposals, R&D contracts, etc. | 17 |
| | Other | Types of data not captured in the GenAI Input Considerations subcodes listed above. | 5 |

**Table 2: Subcodes and Definitions for Data Guidelines for GenAI Use in Research**

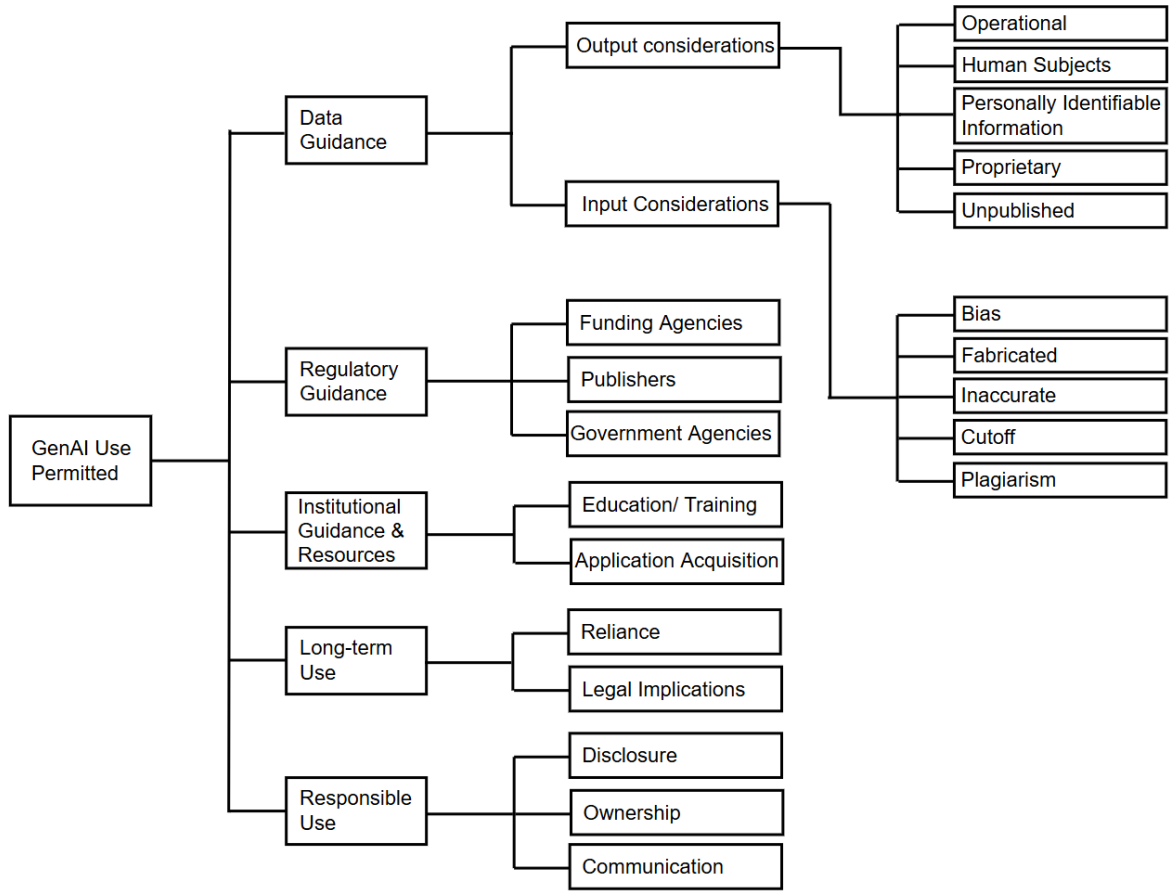

*Figure 2: Inductive framework of coding process – the second layer contains categories, third layer shows codes, fourth layer includes subcodes.*

## 5 Findings

*GenAI Permissible Use*
Most guidelines did not outright prohibit the use of GenAI but rather left the decision to use GenAI up to the researchers (93%, N=28). While the decision to use GenAI was left to researchers' discretion, the guidelines listed both potential benefits and risks associated with GenAI in the research process. A typical directive might read: "If you decide to use generative AI in your research, keep in mind the following items …" [1]. Very few guidelines disallowed the use of GenAI in the research process (7%, N=2) largely on the basis of risks and consequences. For example: one guideline provided updated advisory by highlighting the "prohibited" aspects from federal and publisher organizations [24].

*Data Guidance with the Use of GenAI Tools*
Many of the guidelines mentioned risks and/or implications associated with entering data into GenAI tools or using output from GenAI tools (see Table 2 above). We further examined the policies for guidance related to data output, i.e. what is produced by the GenAI application, and data input, i.e. what information or prompt is entered to get an output. Examples are provided in Table 3.

Most guidelines expressed concerns about GenAI outputs. Under the code "GenAI Output Considerations" (N=27, 90%), we identified subcodes for the following: "Biased Output," "Fabricated Output," "Inaccurate Output," "Knowledge Cutoff Date," "Plagiarism," and "Other." See Table 2 for more details. We found that all guidelines mentioned "Inaccurate Output" (N=27, 100%) as a concern, followed by "Plagiarism" (N=24, 89%), "Biased Output" (N=23, 85%), and "Fabricated Output" (N=16, 59%). A few of the guidelines also referred to last date of data collection to train the system or "Knowledge Cutoff Date" (N=7, 26%) and those that captured other types of output "Other" (N=11, 41%).

Some guidelines elaborated on their concerns about "Fabricated Output" describing a phenomenon known as "hallucinations." GenAI has been known to generate output that is either inaccurate, entirely fabricated, or biased but presented as if it were factual, often referred to as "AI hallucinations." This can introduce risks to researchers when producing unique findings, such as falsified information and inaccurate conclusions (Table 3, (i)). Another ethical issue that was touched on under "Plagiarism" was that GenAI output can be viewed as plagiarism when the output generated is not properly cited or citations are missing completely (Table 3, (ii)). Another aspect considered with regard to GenAI outputs under the code "Knowledge Cutoff Date" is that they may be limited when attempting to answer questions about current events. This is because GenAI models are trained on historical data and may be limited in terms of providing up-to-date information, also known as having a knowledge cutoff date (Table 3, (iii)). Regarding "Other" types of GenAI output mentioned, some of these outputs could include information that presents risks or consequences if used in publishing research (Table 3, (iv)).

Those we coded as "GenAI Input Considerations" (N=26, 87%), emphasized what types of data should not be entered into GenAI tools, as they may violate certain data governance regulations, as well as data privacy concerns. This code included the following subcodes: "Operational Data," "Human Subjects Data," personally identifiable information or "PII," "Proprietary Data," "Unpublished Research," and "Other." See Table 2 above for more detail about the codes.

A majority of guidelines that considered outputs mentioned "PII" (N=18, 69%) and "Unpublished Research" (N=17, 65%); almost half mentioned "Proprietary Data" (N=12, 46%), and about a third mentioned "Human Subjects Data" (N=10, 38%), followed by "Operational Data" (N=8, 31%). In general, guidelines emphasized risks associated with inputting sensitive data into GenAI tools for individuals or organizations if their information was exposed, as well as potential breach of non-public research data. A common concern involved entering PII into GenAI tools, which could inadvertently lead to the exposure of individuals' personal information if GenAI were to be trained using those outputs. This was viewed as a potential violation of various federal and international data privacy laws (e.g., HIPAA, FERPA, GDPR, etc.) (Table 3, (v)).

One guideline emphasized the risk of unpublished research data, arguing that entering this type of data could lead to a breach of confidentiality (Table 3, (vi)). Another set of guidelines expressed concern about entering proprietary data into GenAI tools, saying that this could lead to various legal implications, which are discussed in the Legal Implications with the use of GenAI findings (Table 3, (vii)).

Some guidelines were also concerned with the entry of human subjects data into GenAI tools because it can lead to a similar issue with PII in which supposedly anonymized data can be re-identified, violating

IRB protocols (Table 3, (viii)). Operational data is in terms of data that is used in various administrative functions within a university (e.g., financial data, university data, writing emails, meeting notes, etc.). Some guidelines worried that operational data may be classified as sensitive, especially those that contain discussions surrounding non-public matters (Table 3, (xi)). Regarding "Other" types of data that should not be entered into GenAI tools, this refers to data outside of what is discussed above and could include training data for the development of GenAI tools (Table 3, (x)).

| Example Statements Related to Data Concerns with Using GenAI for Research |
|---|
| **(i)** "It is possible for AI-generated content to be inaccurate, biased, or entirely fabricated (sometimes called 'hallucinations')." [4] |
| **(ii)** "Plagiarism: generative AI can only generate new contents based on, or drawn from, the data that it is trained on. Therefore, there is a likelihood that they will produce outputs that are similar to the training data, even to the point of being regarded as plagiarism if the similarity is too high." [21] |
| **(iii)** "Knowledge Cutoff Date. Many Generative AI models are trained on data up to a specific date and are therefore unaware of any events or information produced beyond that. For example, if a Generative AI is trained on data up to March 2019, they would be unaware of COVID-19 and the impact it had on humanity, or who is the current monarch of Britain. You need to know the cutoff date of the Generative AI model that you use in order to assess what research questions are appropriate for its use." [14] |
| **(iv)** "Please also note that some generative AI tools, such as OpenAI, explicitly forbid their use for certain categories of activity, including harassment, discrimination, and other illegal activities. An example of this can be found in … OpenAI's usage policy document." [17] |
| **(v)** "Ensure that AI systems comply with applicable laws and regulations governing the collection, storage, and use of student data, including the Family Educational Rights and Privacy Act (FERPA), the General Data Protection Regulation (GDPR), Health Insurance Portability and Accountability Act (HIPAA), and other relevant state and federal privacy laws." [10] |
| **(vi)** "Uploading information (e.g., research data, grant proposals, unpublished manuscripts, or analytical results) to a public AI tool is equivalent to releasing it publicly…" [23] |
| **(vii)** "GenAI tools should not be assumed a priori to be private or secure. Users must understand the potential risks associated with inputting sensitive, private, confidential, or proprietary data into these tools, and that doing so may violate legal or contractual requirements, or expectations for privacy." [5] |
| **(viii)** "... inputting interview data to perform preliminary analysis creates the possibility that quotations or other information from research subjects could become public, and potentially, that subjects could also be identified." [4] |
| **(ix)** "For meetings that will involve discussions of a sensitive nature (e.g. personal, confidential, financial, IP, proprietary, personnel, etc.), do not use AI automated meeting tools to record and capture discussions, measure attendee engagement, etc., as the data generated by these tools may be considered Public Records. Be cognizant of virtual meetings where AI meeting tools may be used, inquire with the meeting host about the use of these tools if unsure, and decline participation in the meeting if the host insists on using these tools." [1] |
| **(x)** "Ensuring data privacy and security while maintaining data diversity and representativeness is also a concern. This is also true for AI systems that use unsupervised learning, i.e., that detect patterns in unlabeled data." [20] |

Table 3: Data Related Concerns and Guidance for GenAI Use in Research

*Long-Term Implications of GenAI in Research*
Many guidelines alerted researchers against becoming over reliant on using GenAI systems without developing a good understanding of their limitations or potential for errors (N=25, 83%) and emphasized that they should be cautious when using GenAI tools and verify the output (Table 4, (i)). They emphasized the need for human oversight GenAI tools (Table 4, (ii)) and reminded the researchers that the final responsibility for research rested with them (Table 4, (iii-iv)). The responsible activities for researchers included understanding the legal ramifications of GenAI use, identified in 77% (N=23) of the guidelines, including agreeing to terms and conditions of a GenAI tool, liability to an individual or organization, research misconduct accusations, intellectual property (IP) rights, copyright issues, etc.

| **Example** |
|---|
| **(i)** "Overreliance: The risk of relying excessively on AI systems without considering their limitations or potential errors." [7] |
| **(ii)** "In any use of GAI, there should be self-awareness and recognition that AI is a tool for research. It is a tool of human design and use like any technology. It does not supplant nor surpass human oversight or context of being an independent tool used by humans for its' benefit." [14] |
| **(iii)** "Most public generative AI tools use "clickwrap" or "clickthrough" agreements to get users to accept policies and terms of service before using the tool. Individuals who accept clickthrough agreements without university approval may face personal liability for compliance with the terms and conditions." [2] |
| **(iv)** "AI tools paraphrase from various sources which could result in plagiarism, which would in turn constitute research misconduct or lead to intellectual property issues." [19] |

**Table 4: Long-Term Implications of GenAI Use in Research**

*Regulatory Guidance and Standards for GenAI Use*
A majority of guidelines (N=24, 80%) referred to the regulatory context around the use of GenAI and listed at least one federal, publisher, or other external policy as a resource. This was especially the case when researchers were applying for funding and writing research proposals (Table 5, (i)-(ii)). Similarly, researchers were cautioned about disseminating or publishing their research and asked to refer to specific guidelines from publishers (N=23, 77%). These external guidelines mention some of the themes mentioned throughout this paper, such as responsible use of GenAI, data input/output, legal implications, and more (Table 5, (iii)). The other aspect captured here is reference to broader guidelines (N=9, 30%) released by the White House, National Institute of Standards and Technology (NIST), Executive Orders (EO), etc. These included "White House: "Blueprint for an AI Bill of Rights," "Executive Order on Safe, Secure, and Trustworthy AI," "FACT SHEET: President Biden Issues Executive Order on AI," and "OMB Releases Implementation Guidance Following Executive Order" [26].

| Example |
|---|
| **(i)** "A June 2023 notice from NIH specifically prohibits the use of generative AI tools in grant reviews… This is important because grant applications often contain intellectual property that when shared to a generative AI tool could contribute to future AI output" and "Similarly, a December 2023 notice from NSF states… 'NSF reviewers are prohibited from uploading any content from proposals, review information and related records to non-approved generative AI tools. If reviewers take this action, NSF will consider it a violation of the agency's confidentiality pledge and other applicable laws, regulations and policies'" [3] |
| **(ii)** "Use of GAI in NSF proposals should be indicated in the project description. Specifically, it [Notice to research community: Use of generative artificial intelligence technology in the NSF merit review process] states: 'Proposers are responsible for the accuracy and authenticity of their proposal submission in consideration for merit review, including content developed with the assistance of generative AI tools'" and "Although NIH specifically prohibits GAI in the peer review process, they do not prohibit the use of GAI in grant proposals. They state an author assumes the risk of using an AI tool to help write an application…" [14] |
| **(iii)** "Journals have different rules for reporting the use of generative AI in manuscripts. Generally, journals, including those by the publishing houses of Taylor and Francis and Springer have stated that input from AI must be detailed in the Materials and Methods section, Acknowledgement section, or similar section for transparency. Any publication of reported results should disclose the use of a generative AI tool, which tool, for what parts of the publication and how it was used. Other best practices include indicating the specific language model in addition to the generative AI tool, as well as the date(s) of use, e.g., 'ChatGPT Plus, GPT-4, 19-20 September 2023.'" [19] |

**Table 5: Regulatory Guidance and Standards for GenAI Use**

*Responsible Use of GenAI*

A range of guidelines provide guidance on how to use GenAI in a responsible manner including how to disclose the use of GenAI (N=23, 77%). Some mentioned that GenAI should be listed as an author, while others stated that GenAI could not qualify as an author (Table 6, (i) and (ii)). For the code "Responsible Party for GenAI Usage" (N=20, 67%), the guidelines stated that an individual or party would be held accountable for applications of GenAI in the research process (Table 6, (iii)). About the code "Communicating Responsible Use of GenAI" (N=9, 30%), the guidelines highlighted the importance of lead researchers discussing responsible ways to use GenAI in the research process (Table 6, (iv)).

| Example |
|---|
| **(i)** "Duke researchers employing GenAI tools should always provide attribution. Treat the AI as the author and cite appropriately…" [6] |
| **(ii)** "... use of AI writing tools may be inappropriate (such as generating output that is submitted as one's own original work) or even banned (e.g., listing such a tool as an author [Thorp, 2023] ...)" [22] |
| **(iii)** "Investigators are responsible for maintaining research integrity, rigor, and reproducibility in their work. Undoubtedly, AI will contribute intentional and unintended forms of plagiarism and falsification of data, as have other new technologies. As such, we must be particularly vigilant when employing these new technologies." [30] |
| **(iv)** "Stewardship ~ using resources efficiently attending to one's responsibilities within the scientific enterprise. One of a researcher's responsibilities is living up to the values that keep the research enterprise trustworthy. Regarding the use of AI writing tools, many of the issues are described above. An additional responsibility is being aware of AI writing tools and their potential impacts (positive and negative) on the research enterprise and mentoring the next generation of researchers in their responsible use." [22] |

**Table 6: Responsible Use Guidance for GenAI in Research**

*Institutional Guidance and Resources*

Some guidelines provided institutional-specific resources and procedures to follow with the use of GenAI. The guidelines stated that it is crucial for researchers to stay up-to-date with the latest advancements in GenAI (N=9, 30%) (Table 7, (i)) and a few mentioned that resources such as workshops were available for researchers (Table 7, (ii)). Guidelines also stated specific protocols and procedures for procuring GenAI tools (N=8, 27%) (Table 7, (iii)).

| Example |
|---|
| **(i)** "It is a shared responsibility to stay informed about relevant developments surrounding generative AI… Everyone involved in research should make efforts to stay informed about relevant emerging AI tools, research studies, and ethical guidelines, and should take advantage of professional development opportunities to enhance their AI integration skills." [23] |
| **(ii)** "In order to educate researchers on the use of GenAI, communication and outreach are key. We should educate researchers about the central offices that issue training, guidance, etc. that can help them, rather than leaving them to rely on potentially siloed offices in the units that may not provide consistent advice." [5] |
| **(iii)** "The responsible office, the Division of Digital Learning, shall establish a standardized and transparent approval process for the acquisition, development, and/or deployment of AI technologies. The approval process requires the thorough review and ratification of AI technology by designated authorities in consultation with Procurement Services, Information Technology Services, and other relevant University departments to ensure compliance with University AI policy." [10] |

**Table 7: Institution-Specific Efforts and Resources**

## 6 Analysis of the Overlaps

We conducted an overlap analysis across different categories to (a) identify where institutions are aligning verses taking divergent actions and paths; (b) identify the tradeoffs involved between GenAI use and holistic awareness of the risks; and (c) highlight the overall approach of the institutions including whether the guidelines focus on risk management, rules and compliances or encouraging researchers for innovative use of GenAI technologies. We identified 9 such instances cases (Table 8) and calculated their overlapping percentages (Fig. 3). We found that incorporating all three types of external sources (funding agency, research publishers, and government agency) is not a common practice (~17% of the total) and only a few refer to government agency guidelines. There is a significant overlap in the occurrence of funding and publishing guidelines (70%). Overall, universities are focused on ensuring compliance throughout the research life cycle from funding acquisition to result dissemination but guidelines appear to be more focused on compliance with immediate research standards than broader federal regulations. Almost 64% of the universities that refer to funding and publisher guidelines, also discussed the disclosure of GenAI use. Finally, there is also a significant overlap when it comes to providing guidance on both the data input and output sides of GenAI use (80%). We also uncovered a potential gap connecting responsible parties with guidelines on communicating responsible use. While 66.67% of the institutions seem to mention who will be responsible for ensuring responsible use of GenAI and dealing with legal outcomes, only 23.33% of these have guidance for the party responsible to communicate best practices, ethics, approved uses, benefits, and risks to research staff, including subcontractors. Moreover, relatively low overlap (16.67%) between communicating responsible use and Education/Awareness suggests there is a need to develop a more comprehensive approach to responsible use and education or training that can empower researchers to navigate through responsible use of Gen AI.

| Case | Overlapping codes | Overlap |
|---|---|---|
| 1 | References Funding Agency Guidelines for Use of GenAI<br>References Research Publisher Guidelines for Use of GenAI<br>Reference to Government Agency Guidelines and/or Policies for the Use of GenAI | 5 (16.67%) |
| 2 | References Funding Agency Guidelines for Use of GenAI<br>References Research Publisher Guidelines for Use of GenAI | 21 (70%) |
| 3 | References Funding Agency Guidelines for Use of GenAI<br>Reference to Other Government Agency Guidelines for the Use of GenAI | 6 (20%) |
| 4 | GenAI Output Considerations<br>GenAI Input Considerations | 24 (80%) |
| 5 | GenAI Output Considerations<br>Legal Implications with the use of GenAI | 22 (73.33%) |
| 6 | GenAI Output Considerations<br>Reliance on GenAI | 24 (80%) |
| 7 | References Funding Agency Guidelines for Use of GenAI<br>References Research Publisher Guidelines for Use of GenAI<br>Disclosure of GenAI | 19 (63.33%) |
| 8 | Responsible Party for GenAI Usage<br>Communicating Responsible Use of GenAI | 7 (23.33%) |
| 9 | Communicating Responsible Use of GenAI<br>Education and Awareness for GenAI Use | 5 (16.67%) |

**Table 8: Overlap between different categories**

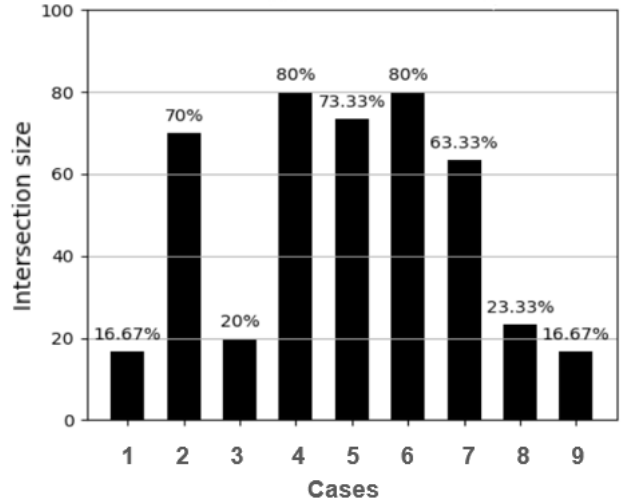

*Figure 3: Overlap between the nine aligning cases*

**7 Discussion**

Analyzing GenAI research guidelines of universities is crucial to understanding the patterns of how universities integrate GenAI into their research practices and more importantly, guidance they are providing to researchers. Overall, institutions appear to be taking a cautious approach towards the use of GenAI in research while being cognizant of associated risks with a significant focus on the data governance, legal implications, and prioritization of risk management. This is not surprising given the uncertainty and novelty of the technology and as a consequence, the guidance places the onus of compliance on individual researchers making their accountable for any lapses, thereby increasing their responsibility.

Specifically, our analysis shows that the specific guidance provided to researchers can be divided into three categories based on the actions that researchers should take (see Fig. 4):
1) *refer* to external sources of information such as funding agencies and publishers to keep updated and use institutional resources for training and education.
2) *understand and learn* about specific GenAI attributes that shape research such as predictive modeling, knowledge cutoff date, data provenance, and model limitations, and about ethical concerns such as authorship, attribution, privacy, and intellectual property issues.
3) *acknowledge and disclose* sources and use of GenAI, communicate effectively about their GenAI use, to mitigate the legal consequences and risks to themselves and their institutions from GenAI use, and *recognize* the long-term implications of reliance on GenAI.

The findings from our study show that the guidance given to researchers touches on different elements put forward by the three frameworks that directed our research process and analysis (Al-fkairy et al, 2024; Lin, 2024; Smith et al, 2024). Consistent with Smith et al., (2024), the guidance refers to both the external and internal contexts i.e., guidelines and policies released by government agencies, funding bodies, and the publishers, and institutional offices that serve as a resources and are relevant for training and education purposes and are ultimately responsible for the implementation of the guidance. Al-fkairy et al., (2024)'s framework is evident in both the technical attributes of GenAI that are relevant, such as training data, modeling, personally identifiable information, and the related ethical concerns that are covered such as attribution, academic integrity, and intellectual property and copyright. Finally, the Lin (2024) framework, similar to Al-kfairy (2024), is reflected in the discussion on ethical concerns with GenAI use.

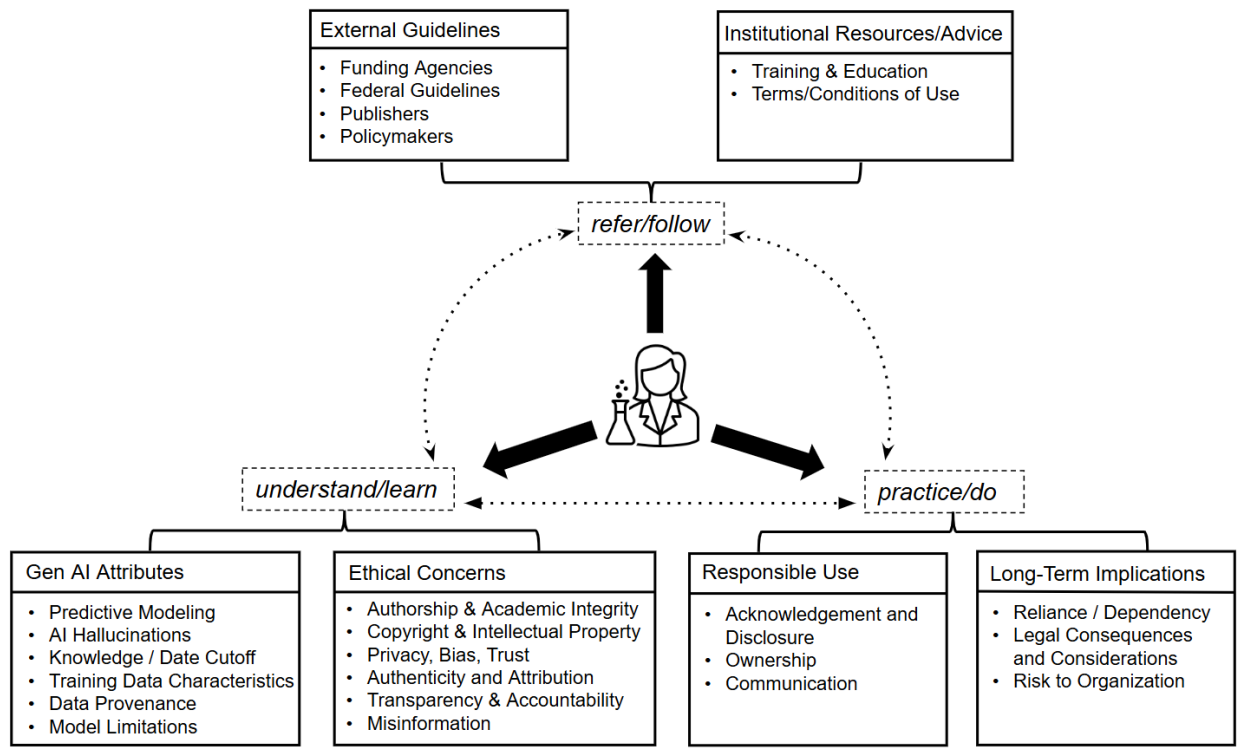

Figure 4: Guidance to Researchers for Using GenAI in Academic Research.

**Implications for Researchers**: From a researcher perspective, we found that as currently framed, the guidelines place a high burden on researchers to learn about and comply with the different rules, and regulations about using GenAI for research. This is not dissimilar from research integrity concerns that researchers have to consider in any case but the complication here, from the perspective of researchers, is lack of clarity around the use of GenAI and an overall lack of transparency related to how technology works. Many developers do not disclose details of how GenAI models are trained, making it difficult for users to understand how to interact with the GenAI tools while attempting to understand how their data will be used. Furthermore, even though researchers are encouraged to disclose GenAI use, there is no established tool or guidance on how to detect the AI generated content (Dalalah & Dalalah, 2023). Moreover, while most guidelines allow guided use, they often lack specific examples and scenarios that would help researchers navigate their daily work. This confusion is especially apparent with authorship issues, which have different guidelines from different actors – funding agencies, federal government, publishers – as each is concerned with a different level of liability. Therefore, for the most part, using LLMs to write is seen as a negative and most academics rather err on being more careful than innovative (Abernethy, 2024). Within the U.S. context, the authorship issue is further complicated by the hegemony of English being the primary language for research and publication and many academics in the U.S. not needed external editorial or writing services. Especially in the humanities and social sciences, writing is considered a component of scholarship itself. Therefore, the issue is seen as controversial and there is hesitancy to use or allow LLMs for writing support. The situation for researchers is further complicated by the responsibility placed on them to understand legal issues embedded within "Terms and Conditions" agreements that are complex, complicated, and often beyond their expertise to evaluate. These "Terms and Conditions" documents obfuscate considerations such as ownership of training data, IP rights, copyright, trademarks, and patents that could lead to liabilities, such as accusations of research misconduct. Multiple lawsuits have been filed against GenAI companies over copyrighted data used for model training, alleging that the defendants violated copyright by training on works for AI models that are then capable of generating outputs that mimic, compete with, or reproduce those works. Institutions that lack guidance on this communication related aspect may be prone to disadvantages of uninformed use of GenAI tools leading to legal implications. From a practical standpoint the guidance reflects the complexity of working within larger policy and institutional contexts. Academic institutions are just one cog within the research enterprise that also includes funding agencies, publishers, and other actors. Therefore, we can see within the HEI guidelines a propensity to refer researchers to external guidelines for many of the concerns. The other practical concern is related to training and education. The gap between the current knowledge about AI and GenAI and what is needed to evaluate their use is extremely high for many academics and it is unclear that it is not a bridge too far to expect them to learn, understand, and make the correct judgments.

**Barriers to Implementation and Other Challenges**: One of the biggest practical concerns with almost no resolution in sight, within the U.S. context at least, is a lack of information about how these applications are trained and how the algorithms work. Even though many efforts are being made towards regulations and guidelines for companies to make their applications more transparent and open to evaluation, that is currently not the case for the most used tools such as ChatGPT. It is also quite unfeasible that companies developing foundational models-based applications will care or do care about academic research to tailor their solutions for this population. It is feasible that industry-directed solutions would come at some stage but given the costs and investments currently required, solutions tailored for academia do not look feasible in the near future. Given these limitations, although a practical approach is the right one, it is unclear how it can be implemented. The low hanging fruit for HEIs are better training programs and hiring experts within research offices that can help academics navigate this landscape. The other feasible approach, as taken by some funding agencies, is to release clear instructions. Although these are often limiting, at least they provide a common ground and are fair. Finally, one aspect of the use of GenAI that is a challenge and referred to by a few guidelines is the trade-off between being too conservative with GenAI use at the expense of innovation. Once again, like other aspects of GenAI use, this decision was also left to the researchers to make. Although the guidance does not directly include researchers' perspective, a limitation also of our work, it is reasonable to assume that many aspect of current guidance are frustrating for the researchers as there are contradictory policies and a large amount of ambiguity. We foresee that as the use of GenAI becomes more stabilized, a more coherent set of guidelines and guidance will be available to researchers.

## 8 Exemplars

Overall, we think there are several exemplary guidelines documents that can form the basis for a more comprehensive approach in the future and also provide a model for articulating concerns and possibilities. These guidelines capture most of the issues identified above, include useful examples from across different stages of research, and have a level of detail and clarity that is useful for researchers.

- Cornell University's [5] guideline document is one of the extensive and comprehensive. It discusses GenAI use at different stages of research including conception and execution, dissemination, translation, funding, and compliance. It also discusses considerations that emphasize specificity along with other important perspectives related to data handling and human subjects. Most of all, this guideline aligns with almost all of the aspects of the exemplary aspects discussed in this article.
- Another exemplary guideline is from Old Dominion University [10]. This guideline encourages researchers to engage critically with AI-generated results. It provides practical strategies to deal with bias and limitations of using GenAI by advocating for human oversight, validation, and communicating responsible use. The guideline document lists specific examples of how AI can be used in different research areas (e.g., data analysis, modeling, simulation). Finally, there are additional useful references to several supporting resources.
- While several university guidelines start with addressing the risks and limitations of GenAI, the guideline from Texas A&M University-College Station [11] starts with mentioning best practices for researchers which are quite forward-thinking and establishes a solid foundation for the responsible use of GenAI. Rather than setting boundaries, the guideline encourages researchers to be collaborative in meeting the legal and ethical expectations. Similarly, the guideline from University of Maine [20] balances the potential of GenAI with practical considerations in research. Along with challenges and risks, it discusses how GenAI can be used across research settings. It also mentions different GenAI tools (LLMs: ChatGPT, Gemini, Image-generators: DALL-E, Midjourney and now Sora for videos) making the guideline versatile for various input and output considerations. The guideline document provides a detailed list of opportunities for using AI and GenAI in research.
- University of Michigan-Ann Arbor [21] addresses concerns related to GenAI use in an easy question-answer format. It mentions all the necessary questions and provided detailed answers to those questions that a researcher might have / need to know before using GenAI in research (e.g. How do I decide which generative AI to use in research?). Their guidelines also highlight their effort for education and training such as a guide for generating, editing, and reviewing code with ChatGPT 4.0, along with a tutorial for coding using local tools like GitHub Copilot.
- Finally, the guideline from Vanderbilt University [27] is a great example of productivity driven and use case-based guidelines on GenAI use. By explaining the use cases this guideline highlighted the significance of understanding field-specific regulations. At the end of the document, there are useful external references such as tips for using GenAI, prompt patterns and prompt engineering for the ChatGPT course.

## 9 Limitations and Future Work

This analysis has been limited in terms of both the dataset and thematic approach used for analysis. We recognize these limitations and recommend that future work be more comprehensive by adopting additional data and methods. For instance, researchers can examine how these policies are being implemented in practice and their impact on research outcome through surveys and interviews with researchers and policymakers. Conducting surveys and interviews with policy makers, researchers, faculty and staff involved in research can provide significant insights into how these guidelines are perceived and applied in the real-life setting and identify potential gaps between the guidelines and the lived experiences of researchers. Evaluating measures such as researcher adoption of GenAI, changes in output quality, and legal consequences faced by the researchers can be helpful to highlight the gaps and opportunities. There is also an opportunity to design and propose a standardized framework which is all inclusive. The analysis showed the institutional guidelines are broad and general. Domain specific guidance and detailed frameworks can be significant and adaptable across different research cycles and disciplines. This can involve proposing modular guidelines that discuss GenAI use and implications at each stage of the research lifecycle. Finally, comparative analysis among university guidelines from different countries can reveal cultural and legal influence. While researchers follow the basic rules in the case of ethics and conducting research, different countries and nations shape how universities approach guiding researchers based on their cultural, legal, and regulatory differences (Gray et al., 2017). The international perspective cannot be validated by the findings from U.S. centric study only. To get a holistic understanding of the global GenAI research policies, a cross-country analysis has the potential to reveal the best and standardized practices.

**Author Contributions**
Author 1 contributed to conceptualization, data analysis, writing, and revision. Author 2 contributed to conceptualization, preliminary analysis, writing, revision, and project fundings. Author 3 contributed to data collection and analysis. Author 4 contributed to conceptualization, writing, and project funding.

**Acknowledgements**
This work is partly supported by US NSF Awards 2319137, 1954556, and USDA/NIFA Award 2021-67021-35329. Any opinions, findings, and conclusions or recommendations expressed in this material are those of the authors and do not necessarily reflect the views of the funding agencies.

student privacy policy documents. *The Journal of Higher Education* 91, 7 (2020), 1149–1178 (2020).

Chawla, D. S. Is ChatGPT corrupting peer review? Telltale words hint at AI use. *Nature*, 628, 483–484 (2024).

Dalalah, D. and Dalalah, O.M., 2023. The false positives and false negatives of generative AI detection tools in education and academic research: The case of ChatGPT. *The International Journal of Management Education*, *21*(2), p.100822.

Delios, A., Tung, R. L. and van Witteloostuijn, A. How to intelligently embrace generative AI: the first guardrails for the use of GenAI in IB research. *Journal of International Business Studies* pp. 1–10 (2024).

de-Lima-Santos, M.F., Yeung, W.N. and Dodds, T. Guiding the way: a comprehensive examination of AI guidelines in global media. *AI & Society*, pp.1-19 (2024).

Duah, J.E. and McGivern, P. How generative artificial intelligence has blurred notions of authorial identity and academic norms in higher education, necessitating clear university usage policies. The *International Journal of Information and Learning Technology*, 41(2), pp.180-193(2024).

Formosa, P., Bankins, S., Matulionyte, R. and Ghasemi, O. Can ChatGPT be an author? Generative AI creative writing assistance and perceptions of authorship, creatorship, responsibility, and disclosure. *AI & Society*, pp.1-13 (2024).

Frank, D., Bernik, A. and Milkovic, M. Efficient generative ai-assisted academic research: Considerations for a research model proposal. IEEE 11th *International Conference on Computational Cybernetics and Cyber-Medical Systems (ICCC)*', IEEE, pp. 000025–000030 (2024).

Godwin, R. C., DeBerry, J. J., Wagener, B. M., Berkowitz, D. E. and Melvin, R. L. Grant drafting support with guided generative ai software. *SoftwareX* 27, 101784 (2024).

Gray, B., Hilder, J., Macdonald, L., Tester, R., Dowell, A. and Stubbe, M. Are research ethics guidelines culturally competent?. *Research Ethics* **13**(1), 23–41 (2017).

Harding, D. and Boyd, P. Generative AI and PhD supervision: A covert third wheel or a seat at the table? In: *ARCOM Conference 2024: Looking Back to Move Forward* 40th Conference, 2-4 September 2024, London (2024).

Hsu, H.-P. Can generative artificial intelligence write an academic journal article? opportunities, challenges, and implications. The Irish Journal of *Technology Enhanced Learning* **7**(2), 158–171 (2023).

Joosten J, Bilgram V, Hahn A, Totzek D. Comparing the ideation quality of humans with generative artificial intelligence. *IEEE Engineering Management Review*. (2024)

**Appendix A: Corpus of R1 Institutions and Relative GenAI Research Guidelines**

| # | R1 Institution | Office/Department | Guideline/Policy Web Link(s) |
|---|---|---|---|
| 1 | Arizona State University | Office of Research | https://researchintegrity.asu.edu/export-controls-and-security/artificial-intelligence |
| 2 | Auburn University | Multiple | https://www.auburn.edu/administration/oacp/AIGuidance.php |
| 3 | Boston College | Office of the Vice Provost for Research | https://www.bc.edu/content/dam/bc1/top-tier/research/VPR/policies/vpr_ai_guidance_1.26.24.pdf |
| 4 | Columbia University in the City of New York | Office of the Provost | https://provost.columbia.edu/content/office-senior-vice-provost/ai-policy |
| 5 | Cornell University | Office of the Vice President for Research | https://research-and-innovation.cornell.edu/generative-ai-in-academic-research/ |
| 6 | Duke University | Office of Research | https://myresearchpath.duke.edu/using-generative-ai-artificial-intelligence-tools-research#faqs |
| 7 | Harvard University | AI-Dedicated Resources | https://www.harvard.edu/ai/research-resources/ |
| 8 | Michigan State University | Office of Research | https://research.msu.edu/generative-ai <br> https://research.msu.edu/generative-ai/guidance |
| 9 | Ohio University-Main Campus | Office of Research | https://www.ohio.edu/research/generative-ai-guidance |
| 10 | Old Dominion University | Multiple | https://ww1.odu.edu/about/policiesandprocedures/bov/bov1200/bov1220 <br> https://www.evms.edu/about_us/ai_resources/specific_usage_guidelines/research_generative_ai_usage_guidelines/ |
| 11 | Texas A&M University-College Station | Office of Research | https://vpr.tamu.edu/memos/best-practices-for-generative-ai-in-research/ <br> https://vpr.tamu.edu/research-resources/resources-on-generative-ai-in-research/ <br> https://vpr.tamu.edu/wp-content/uploads/2024/03/Best-Practices-for-Generative-AI-in-Research-updated-02162447-APPROVED.pdf |
| 12 | The Pennsylvania State University | Multiple | https://ai.psu.edu/guidelines/ <br> https://pennstateoffice365.sharepoint.com/:w:/s/VPR-ORP/EUwbcanBfctBlWG1AG9Cw4EB0fy6TtAbnajHtlNyNiYUUQ?rtime=9oo_sAu13Eg |
| 13 | The University of Alabama | College-Specific | https://as.ua.edu/faculty-resources/ai-statements-of-principle/ |
| 14 | The University of Texas at Arlington | Multiple | https://resources.uta.edu/research/policies-and-procedures/generative-artificial-intelligence.php <br> https://ai.uta.edu/researcher-guidance-for-the-use-of-artificial-intelligence-in-research/ |
| 15 | The University of Texas at San Antonio | AI-Dedicated Resources | https://provost.utsa.edu/academicinnovation/docs/genai_faculty_guide/utsa_faculty-genai-guidelines.1.20.24.pdf |
| 16 | University of Alabama at Birmingham | Office of the President | https://www.uab.edu/ai/guidelines-principles/practical-recommendations<br>https://www.uab.edu/ai/guidelines-principles/generative-ai-and-uab-policy |
| 17 | University of California-Berkeley | Multiple | https://rco.lbl.gov/research-integrity-and-research-ethics/generative-ai-tools-in-research/<br>https://oercs.berkeley.edu/privacy/privacy-resources/appropriate-use-generative-ai-tools |
| 18 | University of Colorado Denver/Anschutz Medical Campus | Office of Research | https://research.ucdenver.edu/resources/use-of-ai-in-pd |
| 19 | University of Kentucky | AI-Dedicated Resources | https://advance.uky.edu/research-recommendations |
| 20 | University of Maine | AI-Dedicated Resources | https://docs.google.com/document/d/1LEb1z8u9WiWe_za18M2WG_RM9TRaBJ-P/edit |
| 21 | University of Michigan-Ann Arbor | AI-Dedicated Resources | https://midas.umich.edu/generative-ai-user-guide/#additional-reading |
| 22 | University of New Hampshire-Main Campus | Other | https://www.unh.edu/teaching-learning-resource-hub/sites/default/files/media/2023-11/unh-guide-to-using-generative-ai-writing-oct-31-2029.pdf |
| 23 | University of North Carolina at Chapel Hill | Office of the Provost | https://provost.unc.edu/generative-ai-usage-guidance-for-the-research-community/ |
| 24 | University of Rochester | Office of the Vice President for Research | https://www.rochester.edu/university-research/updated-advisory-guidance-on-the-use-of-generative-ai-in-research/ |
| 25 | University of Utah | Office of Research | https://integrity.research.utah.edu/ai-research-statement.php |
| 26 | Utah State University | Office of Research | https://research.usu.edu/compliance/ai-in-research |
| 27 | Vanderbilt University | Multiple | https://www.vanderbilt.edu/generative-ai/research/<br>https://www.vumc.org/dbmi/GenerativeAI |
| 28 | Virginia Polytechnic Institute and State University | Office of Research | https://www.research.vt.edu/research-support/forms-guidance/sirc/guidance-using-artificial-intelligence-during-research-activities.html |
| 29 | Washington State University | Office of Research | https://research.wsu.edu/guidelines-policies/generative-ai |
| 30 | Washington University in St. Louis | Office of Vice Chancellor for Research | https://research.wustl.edu/announcements/message-from-the-vice-chancellor-for-research-regarding-artificial-intelligence/ |